\documentclass[aps,prd,twocolumn,groupedaddress,showpacs,showkeys]{revtex4-1}
\usepackage{graphicx}
\usepackage{dcolumn}
\usepackage{bm}
\usepackage{amssymb,amsmath,latexsym,amsfonts}
\begin{document}


\title{Modified Bosonic Gas Trapped in a Generic 3--dim Power Law Potential}

\author{E. Castellanos}\email{elias@zarm.uni-bremen.de}\affiliation{ZARM,
Universitaet Bremen, Am Fallturm, 28359 Bremen, Germany}

\author{C. Laemmerzahl}
\email{claus.laemmerzahl@zarm.uni-bremen.de} \affiliation{ZARM,
Universitaet Bremen,
Am Fallturm, 28359 Bremen, Germany}


\begin{abstract}

We analyze the consequences caused by an anomalous single-particle dispersion relation suggested in several quantum-gravity models, upon the thermodynamics of a Bose--Einstein condensate trapped in a generic 3-dimensional power-law potential. We prove that the condensation temperature is shifted as a consequence of such deformation and show that this fact could be used to provide bounds on the deformation parameters. Additionally, we show that the shift in the condensation temperature, described as a non-trivial function of the number of particles and the trap parameters, could be used as a criterion to analyze the effects caused by a deformed dispersion relation in weakly interacting systems and also in finite size systems.\end{abstract}

\pacs{04.60.Bc, 05.30.Jp, 03.75.Hh}

 \maketitle

\section{Introduction}

The possibility of a deformation in the dispersion relation of microscopic particles, appears in connection with the quest for a quantum theory of gravity \cite{ami,Giovanni1,Claus,Claus1,Kostelecky,amelino1,5,12,13}. This entails, in some schemes, that a possible spacetime quantization has as a consequence a modification of the classical--spacetime dispersion relation between energy $E$ and (modulus of) momentum $p$ of a microscopic particle with mass m \cite{Giovanni1,Claus,Kostelecky}. A deformed dispersion relation
emerges as an adequate tool in the search for phenomenological
consequences caused by this type of quantum gravity models.
Nevertheless, the principal difficulty in the search of
quantum gravity manifestations in our low energy world, is the smallness in the predicted
effects \cite{Kostelecky,amelino1}. If this kind of deformations is
characterized, for instance, by some Planck scale, then the quantum gravity effects
become very small \cite{Giovanni1,Claus}. In the non--relativistic
limit, the deformed dispersion relation can be expressed as follows
\cite{Claus,Claus1}
\begin{equation}
E \simeq
m+\frac{p^{2}}{2m}+\frac{1}{2M_{p}}\Bigl(\xi_{1}mp+\xi_{2}p^{2}+\xi_{3}\frac{p^{3}}{m}\Bigr),
\label{ddr}
\end{equation}
in units where the speed of light
$c=1$, with $M_{p}\simeq 1.2\times 10^{28}eV$ the Planck
mass. The three parameters $\xi_{1}$, $\xi_{2}$, and $\xi_{3}$, are
model dependent \cite{Giovanni1,Claus}, and should take positive or
negative values close to $1$. There is some evidence within the
formalism of Loop quantum gravity \cite{Claus,Claus1,5,12} that indicates non--zero values for the three parameters, $\xi_{1},\, \xi_{2},\,
\xi_{3}$, and particulary that produces a
linear--momentum term in the non--relativistic limit \cite{5,13}. 
Unfortunately, as it is usual in quantum gravity phenomenology, the possible bounds associated with the deformation parameters, open a wide range of possible magnitudes, which is translated to a significant challenge.

On the other hand, the use of  Bose--Einstein
condensates, as a possible tool in the search of quantum--gravity
manifestations (for instance, in the context of Lorentz violation or
to provide phenomenological constrains on Planck--scale physics) has produced an enormous amount of interesting publications
\cite{Colladay,Donald,Camacho,CastellanosCamacho,CastellanosCamacho1,CastellanosClaus,r1,r2,echa,Eli1}.
It turns out to be rather exciting to look for the effects in
the thermodynamic properties associated with Bose--Einstein
condensates caused by the quantum structure of space--time. 

In a previous report \cite{CastellanosClaus}, we were able to prove, that the condensation temperature of the ideal bosonic gas, is corrected as a consequence of the deformation in the dispersion relation. Moreover, this correction described as a non--trivial function of the number of particles and the shape associated with the corresponding trap it could provide representative bounds for the deformation parameter $\xi_{1}$. We have proved that the deformation parameter $\xi_{1}$ can be bounded, under typical conditions, from $|\xi_{1}|\lesssim10^{6}$ to $|\xi_{1}|\lesssim10^{2}$, by using different classes of trapping potentials in the thermodynamic limit. In the case of a harmonic oscillator--type potential, we have obtained a bound up to  $|\xi_{1}|\lesssim10^{4}$.  In references \cite{Claus,Claus1} it was suggested
the use of ultra--precise cold--atom--recoil experiments to constrain
the form of the energy-momentum dispersion relation in the
non--relativistic limit. There, the bound
associated with $\xi_{1}$ is at least, four orders of magnitude smaller
than the bound associated with a Bose--Einstein condensate trapped in a harmonic oscillator in the ideal case obtained in \cite{CastellanosClaus}.
 
In a more realistic system, finite size effects and interactions among the constituents of the gas must be taken into account. To this aim, let us propose a particularly simple modified Hartree-Fock type
spectrum, in the semi--classical approximation, which basically consist in the
assumption that the constituents of the gas behave like non--interacting bosons moving in a self--consistent mean field, valid when the semiclassical energy spectrum $\epsilon_{p}$ is bigger than the associated chemical potential $\mu$, for dilute gases \cite{Dalfovo,Pethick}
\begin{equation}
\label{HF0} \epsilon_{p}=\frac{p^{2}}{2m}+ \alpha
p+U(\vec{r})+2U_{0}n(\vec{r}),
\end{equation}
where $p$ is the momentum, $m$ is the mass of the particle, and the
term $\alpha p$, with $\alpha=\xi_{1}\frac{mc}{2M_{p}}$ in ordinary
units, is the leading order modification in expression (\ref{ddr}),
with $c$ the speed of light. The term $2U_{0}n(\vec{r})$ is a mean field generated by the
interactions with the other constituents of the bosonic gas, being $n(\vec{r})$ the spatial density of the cloud \cite{Dalfovo}. The coupling constant $U_{0}$ is related
to the s--wave scattering length $a$ through the following expression
\begin{equation}
U_{0}=\frac{4 \pi \hbar^{2}}{m} a.
\end{equation}
The potential term
\begin{equation}
\label{potgen}
 U(\vec{r})=\sum_{i=1} ^ {d} A_{i}\Bigg|\frac{r_{i}}{a_{i}}\Bigg|^{s_{i}},
\end{equation}
is the generic 3--dimensional
power--law potential, where $ A_{i}$ and $a_{i}$ are energy and
length scales associated with the trap \cite{Jaouadi}. Additionally,
$r_{i}$ are the $d$ radial coordinates in the $n_{i}$--dimensional
subspace of the 3--dimensional space. The sub--dimensions $n_{i}$
satisfy the following expression in three spatial dimensions
\begin{equation}
\sum_{i=1} ^ {d} n_{i}=3. \label{subdim}
\end{equation}
If $d=3$, $n_{1}=n_{2}=n_{3}=1$, then the
potential becomes in the Cartesian trap. If $d=2$, $n_{1}=2$ and
$n_{2}=1$, then we obtain the cylindrical trap. If $d=1$, $n_{1}=3$,
then we have the spherical trap. If $s_{i}\rightarrow\infty$, we
have a free gas in a box. The external potential included in (\ref{HF0}) is quite general. Different
combinations of these parameters give different classes of
potentials, according to (\ref{potgen}). It is noteworthy to mention that the use of these generic potentials, opens the possibility to adiabatically cool the system in a reversible way, by changing the shape of the trap \cite{Dalfovo}.
The analysis of a Bose--Einstein condensate in
the ideal case, weakly interacting, and with a finite number of
particles, trapped in different potentials show that the main
properties associated with the condensate, and in particular the
condensation temperature, strongly depends on the trapping potential under consideration  \cite{Jaouadi,bagnato,yukalov,yukalov1,yukalov2,grossmann,Giorgini,ketterle,Haugerud,Li,zijun,Salasnich,Zobay,ET}.
Additionally, the characteristics of the potential (in particular, the parameter
that defines the shape of the potential) has a strong impact on the
dependence of the condensation temperature with the number of particles (or the associated density).

The main goal of this work is to analyze the shift in the condensation temperature caused by a deformed dispersion relation in weakly interacting systems and also in systems containing a finite number of particles. We stress that these systems could be used, in principle, to obtain criteria of viability for possible signals coming from Planck scale regime, by analyzing some relevant thermodynamic variables, for instance, the number of particles, and the frequency associated with the trap, when $|\xi_{1}|\lesssim 1$.
\section{Condensation Temperature in the Thermodynamic Limit; $U_{0}=0$}

 Due to an extensive use of some results, let us briefly summarize the results obtained in \cite{CastellanosClaus}.  From (\ref{HF0}), the case $U_{0}=0$ leads to
\begin{equation}
\label{HFI} \epsilon_{p}=\frac{p^{2}}{2m}+ \alpha
p+U(\vec{r}).
\end{equation}
In the semiclassical approximation, the single--particle
phase--space distribution may be written as \cite{Dalfovo,Pethick}
\begin{equation}
\label{SPSD}
n(\vec{r},\vec{p})=\frac{1}{e^{\beta(\epsilon_{p}-\mu)}-1},
\end{equation}
where $\beta=1/\kappa T$, $\kappa$ is the Boltzmann constant, $T$ is the temperature, and $\mu$ is the chemical potential.
The number of particles in the 3--dimensional space obeys the
normalization condition \cite{Dalfovo,Pethick},
\begin{equation}
 N=\frac{1}{(2 \pi \hbar)^{3} }\int d^{3}\vec{r}\hspace{0.1cm} d^{3} \vec{p}\hspace{0.1cm} n(\vec{r},\vec{p}),
\label{NC}
\end{equation}
where
\begin{equation}
\label{n1} n(\vec{r})=\int  d^{3} \vec{p} \hspace{0.1cm}
n(\vec{r},\vec{p}),
\end{equation}
is the spatial density. Using expression
(\ref{HFI}), and integrating expression (\ref{SPSD}) over momentum, with the help of (\ref{n1}), we get the spatial
distribution associated with our modified semi--classical spectrum
(\ref{HFI})
\begin{eqnarray}
\label{MSD10}
 n(\vec{r})&=&\lambda^{-3} g_{3/2}\Bigl(e^{\beta(\mu_{eff}-U(\vec{r}))}\Bigr)\\\nonumber &-&
\alpha \lambda^{-2}\Bigl(\frac{m}{\pi \hbar}\Bigr)
g_{1}\Bigl(e^{\beta(\mu_{eff}-U(\vec{r}))}\Bigr)
\\\nonumber&+&\alpha^{2}\lambda^{-1}\Bigl(\frac{m^{2}}{2\pi
\hbar^{2}}\Bigr)g_{1/2}\Bigl(e^{\beta(\mu_{eff}-U(\vec{r}))}\Bigr)
\end{eqnarray}
where $\lambda=\Bigl(\frac{2 \pi \hbar^{2}}{m \kappa
T}\Bigr)^{1/2}$, is the de Broglie thermal wavelength,
$\mu_{eff}=\mu+m\alpha^{2}/2$ is an effective chemical potential,
and $g_{\nu}(z)$ is the so--called Bose--Einstein function defined
by \cite{Phatria}
\begin{equation}
g_{\nu}(z)=\frac{1}{\Gamma(\nu)}\int_{0}^{\infty}\frac{x^{\nu-
1}dx}{z^{-1}e^{x}-1}. \label{BEF}
\end{equation}
If we set $\alpha=0$ in equation (\ref{MSD10}) we recover the usual
result for the spatial density in the semiclassical approximation
\cite{Dalfovo,Pethick}.
 By using the properties of the Bose--Einstein functions \cite{Phatria}, assuming that $m\alpha^{2}/2 << \kappa T$ and integrating the normalization condition (\ref{NC}),
we obtain an expression for the number
of particles $N$ to first order in $\alpha$
\begin{eqnarray}
\label{NPINT10}
N-N_{0}&=&C\Pi_{l=1}^{d}A_{l}^{-\frac{n_{l}}{s_{l}}}a_{l}^{n_{l}}\Gamma
\Bigl(\frac{n_{l}}{s_{l}}+1\Bigr)\\\nonumber&\times&\Bigg[\Bigl(\frac{m}{2\pi
\hbar^{2}}\Bigr)^{3/2} g_{\gamma}(z)(\kappa T)^{\gamma}\\\nonumber&-&\alpha
\Bigl(\frac{m^2}{2\pi^{2} \hbar^{3}}\Bigr)g_{\gamma-1/2}(z)(\kappa
T)^{\gamma-1/2}\Bigg],
\end{eqnarray}
where
\begin{equation}
\label{parametergamma}
\gamma=\frac{3}{2}+\sum_{l=1}^{d}\frac{n_{l}}{s_{l}},
\end{equation}
is the parameter that defines the shape of the potential
(\ref{potgen}). In (\ref{NPINT10}), $N_{0}$ are the particles in the ground
state, $\Gamma(y)$ is the Gamma function, and $C$ is a constant
associated with the potential in question. In the case of Cartesian
traps, and in consequence, for a three dimensional harmonic
oscillator potential $\gamma=3$ and $C=8$. If we set $\alpha=0$ in (\ref{NPINT10}) then, we
recover the result given in \cite{Jaouadi}.
 In the thermodynamic limit the conditions for condensation are given by $\mu=0$ and
$N_{0}=0$, which implies hat the Bose--Einstein functions become the corresponding Riemann Zeta functions $\zeta(x)$ \cite{Phatria}. Thus, expression (\ref{NPINT10}) at the condensation temperature is given by
\begin{eqnarray}
\label{temcrit}
N&=&C\Pi_{l=1}^{d}A_{l}^{-\frac{n_{l}}{s_{l}}}a_{l}^{n_{l}}\Gamma
\Bigl(\frac{n_{l}}{s_{l}}+1\Bigr)\Bigg[\Bigl(\frac{m}{2\pi
\hbar^{2}}\Bigr)^{3/2} \zeta (\gamma)(\kappa T_{c})^{\gamma}\,\,\,\,\,\,\,\,\,\,\ \\\nonumber&-&\alpha
\Bigl(\frac{m^2}{2\pi^{2} \hbar^{3}}\Bigr)\zeta(\gamma-1/2)(\kappa
T_{c})^{\gamma-\frac{1}{2}}\Bigg],
\end{eqnarray}
where $T_{c}$ is the condensation temperature. If we set $\alpha=0$ in (\ref{temcrit}), we recover the
usual expression for the condensation temperature $T_{0}$ for a gas
trapped in a generic 3--dim power--law potential in the
thermodynamic limit \cite{Jaouadi}
\begin{equation}
\label{criter}
T_{0}=\Bigg[\frac{N\Pi_{l=1}^{d}A_{l}^{\frac{n_{l}}{s_{l}}}a_{l}^{-n_{l}}}{C\Pi_{l=1}^{d}\Gamma
\Bigl(\frac{n_{l}}{s_{l}}+1\Bigr)}\Bigl(\frac{2 \pi
\hbar^{2}}{m}\Bigr)^{3/2}\Bigg]^{1/\gamma}\frac{1}{\kappa}.
\end{equation}
Now, let us define
\begin{equation}
\label{volumen}
V_{char}=\frac{\Pi_{l=1}^{d}A_{l}^{\frac{n_{l}}{s_{l}}}a_{l}^{-n_{l}}}{C\Pi_{l=1}^{d}\Gamma
\Bigl(\frac{n_{l}}{s_{l}}+1\Bigr)},
\end{equation}
as the characteristic volume associated with the system. We notice that if
$s_{i}\rightarrow\infty$ then, $ V_{char}$ becomes the volume
associated with a free gas in a box (in fact the inverse of the
volume with this definition). In this sense, $V_{char}$ can be
interpreted as the available volume occupied by the gas
\cite{zijun,yukalov1}. At this point, it is noteworthy to mention that
the most general definition of thermodynamic limit can be expressed
as $N\rightarrow \infty,\hspace{0.5cm}V_{char}\rightarrow0$
keeping the product $NV_{char}$ constant, and is valid for
all power law potentials in any spatial dimensionality
\cite{yukalov1}. With the criterion given above, the condensation
temperature in the thermodynamic limit is well defined.
Finally, we can express the shift in the condensation temperature as a function
of the number of particles $N$
\begin{equation}
\label{CTNP} \frac{T_{c}-T_{0}}{T_{0}}\equiv\frac{\Delta
T_{c}}{T_{0}}\simeq\alpha \Omega N^{-1/2\gamma},
\end{equation}
where
\begin{equation}\label{omi}
\Omega=\Bigg(\frac{2m}{\pi}\Bigg)^{1/2}\frac{\zeta(\gamma-1/2)}{\gamma\zeta(\gamma)}
\Bigg(\frac{V_{char}(2\hbar^{2})^{3/2}}{\zeta(\gamma)}\Bigg)^{-1/2\gamma}.
\end{equation}

For the sake of simplicity, let us analyze the case of
spherical traps, where the corresponding potential is given
by $U(r)=A_{1}(\frac{r}{a_{1}})^{s_{1}}$, setting $A_{1}=\hbar
\omega_{0}/2$ and $a_{1}=\sqrt{\hbar/m \omega_{0}}$.
In this case, the shift in the condensation temperature is given by
\begin{equation}
\label{esferico} \frac{\Delta T_{c}}{T_{0}}\simeq\alpha \Omega_{s_{1}}
N^{-s_{1}/3(s_{1}+2)}.
\end{equation}
For different values of $s_{1}$ we obtain, $\frac{\Delta
T_{c}}{T_{0}}\sim\alpha N^{-1/9}$ , for $s_{1}=1$, which corresponds
to a linear trap. For $s_{1}=2$, $\frac{\Delta
T_{c}}{T_{0}}\sim\alpha N^{-1/6}$, which is an isotropic harmonic
oscillator. For  $s_{1}=3$,  $\frac{\Delta T_{c}}{T_{0}}\sim\alpha
N^{-1/5}$. For  $s_{1}=6$,  $\frac{\Delta T_{c}}{T_{0}}\sim\alpha
N^{-1/4}$, and so on. We notice immediately that if
$s_{1}\rightarrow \infty$, after some algebraic manipulation, we are able to obtain the limiting
case of a bosonic gas trapped in a box
\begin{equation}
\frac{\Delta
T_{c}}{T_{0}}\simeq\alpha\frac{2m(V\zeta(3))^{1/3}}{3\hbar}N^{-1/3}.
\end{equation}
Additionally, the current high precision experiments in the case of
$^{39}_{19} K$, with a mass $15 \times 10^{-26}\,Kg$, the shift in the condensation temperature
respect to the ideal result, caused by the interactions among the
constituents of the gas is about $5\times 10^{-2}$ with a $1\%$  of
error \cite{RP}. 
These facts allows us to obtain a bound for the
deformation parameter $|\xi_{1}|\lesssim10^{6}$ for the linear trap $s_{1}=1$
(with frequencies $\omega_{0}/2\pi \sim 10\, Hz$ and $N\sim
10^{6}$), to $|\xi_{1}|\lesssim10^{2}$ corresponding to a free gas in a box $s_{1}\rightarrow \infty$, with
densities about $10^{13}-10^{15}$ \cite{Dalfovo}.
In fact, these bounds could be improved in a system
containing massive bosons and/or lower frequencies but,
where the thermodynamic limit is still valid, trapped in potentials where
the parameter $s_{1}$ is sufficiently large.
Here one important fact is that we are able to improve the bound associated to the
deformation parameter $\xi_{1}$ by the use of different classes of
potentials and it is straightforward to generalize this result to more general
traps by using (\ref{CTNP}).

From (\ref{CTNP}) and (\ref{omi}), we notice that correction in the condensation
temperature depends strongly on the functional form between the
number of particles and the parameters associated to the potential
in question. 
Finally, in the case of a harmonic oscillator potential we obtain, $\frac{\Delta T_{c}}{T_{0}}\sim \alpha N^{-1/6}$ corresponding to a shift of order $\xi_{1}\,10^{-6}$, which allows us to bound the parameter $\xi_{1}$ up to $|\xi_{1}|\lesssim10^{4}$, under typical conditions.

\section{Weakly Interacting Modified Bosonic Gas}

Let us start with the modified Hartree--Fock spectrum (\ref{HF0}).
After some calculations, similar to the previous section, we obtain the spatial density associated to the weakly interacting case
\begin{eqnarray}
\label{MSD}
n(\vec{r})&=&\lambda^{-3}
g_{3/2}\Bigl(e^{\beta(\mu_{eff}-U(\vec{r})-2U_{0}n(\vec{r}))}\Bigr)
\\\nonumber&-&\alpha \lambda^{-2}\Bigl(\frac{m}{\pi \hbar}\Bigr)
g_{1}\Bigl(e^{\beta(\mu_{eff}-U(\vec{r})-2U_{0}n(\vec{r}))}\Bigr).
\end{eqnarray}
If we set $\alpha=0$ in equation (\ref{MSD}) we recover the usual
result for the spatial denstity in the semiclassical approximation
\cite{Dalfovo,Pethick}.
By using the properties of the Bose--Einstein functions \cite{Phatria}, we are able to expand expression (\ref{MSD}) around $U_{0}=0$,
with the result
\begin{eqnarray}
\label{MSD1} n(\vec{r})&\approx& n_{0}(\vec{r})+U_{0}(2\kappa
T)^{-1}\lambda^{-6}\Bigl[g_{3/2}(Z)g_{1/2}(Z)\Bigr]\\\nonumber&+&\alpha
U_{0}(2\kappa T)^{-1}\lambda^{-5}\Bigl(\frac{m}{\pi
\hbar}\Bigr)\Bigl[g_{3/2}(Z)g_{0}(Z)\\\nonumber&+&g_{1}(Z)g_{1/2}(Z)\Bigr],
\end{eqnarray}
where
\begin{equation}
 Z=e^{\beta(\mu_{eff}-U(\vec{r}))},
\end{equation}
being $n_{0}(\vec{r})$ the space density distribution for the ideal
case $U_{0}=0$,
\begin{equation}
n_{0}(\vec{r})=\lambda^{-3} g_{3/2}(Z)-\alpha
\lambda^{-2}\Bigl(\frac{m}{\pi \hbar}\Bigr) g_{1}(Z). \label{n0}
\end{equation}
Integrating the normalization condition (\ref{NC}) and using expression
(\ref{MSD1}) with the corresponding potential (\ref{potgen}), allows
us to obtain an expression for the number of particles as a function
of the chemical potential $\mu$, the temperature $T$, the coupling
constant $U_{0}$, and the deformation parameter $\alpha$
\begin{eqnarray}
\label{NPINT} NV_{char} &=& \Bigg[\Bigl(\frac{m}{2\pi
\hbar^{2}}\Bigr)^{3/2}g_{\gamma}(z_{eff})(\kappa T)^{\gamma}
\\\nonumber&-& \alpha \Bigl(\frac{m^2}{2\pi^{2} \hbar^{3}}\Bigr)g_{\gamma}(z_{eff})(\kappa T)^{\gamma-1/2}
\\\nonumber&-&U_{0} \Bigl(\frac{m}{2 \pi \hbar^{2}}\Bigr)^{3}
G_{3/2,1/2,\gamma-3/2}(z_{eff})(\kappa T)^{\gamma-3/2}
\\\nonumber&+& \alpha U_{0} \Bigl(\frac{m}{2 \pi \hbar^{2}}\Bigr)^{5/2}
\Bigl(\frac{m}{\pi \hbar}\Bigr)(\kappa
T)^{\gamma}\\\nonumber&\times&\Bigl( G_{3/2,0,\gamma-3/2}(z_{eff})
+G_{1,1/2,\gamma-3/2}(z_{eff})\Bigr)\Bigg],
\end{eqnarray}
where
\begin{equation}
G_{\eta,\sigma,\gamma-3/2}(z_{eff})=\sum_{i
j=1}^{\infty}\frac{z_{eff}^{(i+j)}}{i^{\eta}j^{\sigma}(i+j)^{\gamma-3/2}},
\end{equation}
and we have defined an effective fugacity
\begin{equation}
\label{zef}
z_{eff}=e^{\beta(\mu +m \alpha ^{2}/2)}.
\end{equation}
In order to obtain the leading correction on the condensation temperature caused
by the interactions in our deformed
bosonic gas, let us expand (\ref{NPINT}) to first
order in $T=T_{0}$, $\mu=0$, $U_{0}=0$, and $\alpha=0$. Recalling that $T_{0}$ is the condensation temperature in the thermodynamic limit
given by expression (\ref{criter}), with the result
\begin{eqnarray}
\label{NPINT1} NV_{char} &=&\Bigg[\Bigl(\frac{m} {2 \pi
\hbar^{2}}\Bigr)^{3/2}\zeta(\gamma)(\kappa T_{0})^{\gamma}
\\\nonumber&+&[T-T_{0}]\Bigl(\frac{m}{2 \pi \hbar^{2}}\Bigr)^{3/2}
\gamma\zeta(\gamma)\kappa(\kappa T_{0})^{\gamma-1}
\\\nonumber&-&U_{0}\Bigl(\frac{m}{2 \pi \hbar^{2}}\Bigr)^{3}
G_{3/2,1/2,\gamma-3/2}(1)(\kappa T_{0})^{\gamma+1/2}
\\\nonumber&+&\mu\Bigl(\frac{m}{2 \pi \hbar^{2}}\Bigr)^{3/2}
\zeta(\gamma-1) (\kappa T_{0})^{\gamma-1}\\\nonumber&-&\alpha
\frac{m^{2}}{\pi^{2}\hbar^{3}} \zeta(\gamma-1/2)(\kappa
T_{0})^{\gamma-1/2}\Bigg].
\end{eqnarray}
At the condensation temperature $T_{c}$ for large $N$, in the mean field
approach the chemical potential takes the value $\mu_{c}=2\,U_{0}n_{0}(\vec{r}=\vec{0})$ \cite{Dalfovo,Pethick}. In the usual case $\alpha=0$, $n_{0}(\vec{r}=\vec{0})=\lambda_{c}^{-3}\zeta(3/2)$ in
the large $N$ limit, which means that the critical density at the center of the trap is
the same as that of the uniform model \cite{Dalfovo}. However, in our case, we have to modified the value of $\mu_{c}$ at the condensation temperature
according to expression (\ref{n0}), due to the divergent
behavior of the Bose-Einstein functions related to
$n_{0}(\vec{r}=\vec{0})$. When the integrals associated with the Bose--Einstein functions converges,
the value $m\alpha^{2}/2$ is negligible and can be replaced by zero.
Nevertheless, when the integral associated to the Bose--Einstein functions
can diverge at $Z\rightarrow1$ the minimum of the energy associated with
the  system must be taken into account \cite{yukalov1}. In this section we are
interested in the corrections due to $\alpha$ in the large $N$
limit, so we will take as the minimum of energy in the system
$m\alpha^{2}/2$.
Let us define $n_{0}(\vec{r}=\vec{0})$ at the condensation temperature
using expression (\ref{n0}) as follows
\begin{eqnarray}
\label{newn0}
n_{0}(\vec{r}=\vec{0})&=&\lambda_{c}^{-3}g_{3/2}(e^{\beta_{c} m
\alpha^{2}/2})\\\nonumber&-&2\alpha
U_{0}\lambda_{c}^{-2}\Bigl(\frac{m}{\pi
\hbar}\Bigr)g_{1}(e^{\beta_{c} m \alpha^{2}/2}).
\end{eqnarray}
We can define the Bose--Einstein functions 
$g_{3/2}(e^{\beta_{c}m\alpha^{2}/2})$ and
$g_{1}(e^{\beta_{c}m\alpha^{2}/2})$, when $(\beta_{c} m
\alpha^{2}/2) \rightarrow 0$ as \cite{Phatria}
\begin{equation}
\label{g1}
g_{3/2}(e^{\beta_{c}m\alpha^{2}/2})\simeq\zeta(3/2)+\Gamma(-1/2)\Bigl(\frac{m
\alpha^{2}}{2 \kappa T_{c}}\Bigr)^{1/2}
\end{equation}
\begin{equation}
\label{g2}
g_{1}(e^{\beta_{c}m\alpha^{2}/2})\simeq\ln\Bigl(\frac{2\kappa
T_{c}}{m \alpha^{2}}\Bigr).
\end{equation}
Neglecting second order terms in $U_{0}$ and $\alpha$, this allows us
to write $\mu_{c}$ using expression (\ref{newn0}) as follows
\begin{equation}
\label{MUCR} \mu_{c}\simeq 2U_{0}\lambda_{c}^{-3}\zeta(3/2)-2\alpha
U_{0}\lambda_{c}^{-2}\Bigl(\frac{m}{\pi\hbar}\Bigr)\ln\Bigl(\frac{2\kappa
T_{c}}{m \alpha^{2}}\Bigr).
\end{equation}
If we take the limit $\alpha \rightarrow\ 0$, then we recover the usual value for $\mu_{c}$ at the
condensation temperature \cite{Dalfovo,Pethick}. Inserting  (\ref{MUCR}) in (\ref{NPINT1}), we finally obtain 
the shift in the condensation temperature in function of the number of particles
\begin{eqnarray}
\label{CRTI}
\frac{\Delta
T_{c}}{T_{0}}&\simeq&-(aR_{0})^{\frac{1}{2\gamma}}\Bigl(\frac{m\Lambda^{2}}{2\pi
\hbar^{2}}\Bigr)^{\frac{1}{2}}
N^{\frac{1}{2\gamma}}\\\nonumber&+&\alpha\frac{(8 \pi
m)^{1/2}\zeta(\gamma-1/2)} {\zeta(\gamma)\gamma}
(R_{0}N)^{-\frac{1}{2\gamma}}
\\\nonumber&+&\alpha a \frac{4m \zeta(\gamma-1)}{\pi
\hbar\zeta(\gamma)\gamma}\ln
\Bigg(\frac{(R_{0}N)^{\frac{1}{\gamma}}}{m \alpha^{2}}\Bigg),
\end{eqnarray}
where
\begin{equation}
\Lambda=\frac{2\zeta(3/2)\zeta(\gamma-1)
-G_{3/2,1/2,\gamma-3/2}(1)}{\zeta(\gamma) \gamma},
\end{equation}
\begin{equation}
\label{R0} R_{0}=\Bigl(\frac{2 \pi
\hbar^{2}}{m}\Bigr)^{3/2}\Bigg[\frac{V_{char}}
{\zeta(\gamma)}\Bigg].
\end{equation}
Setting $\alpha=0$ in equation (\ref{CRTI}) we recover the usual
shift on the condensation temperature caused by weakly interactions.
Let us analyze the case of spherical traps $U(r)=A_{1}(\frac{r}{a_{1}})^{s_{1}}$, together with $A_{1}=\hbar \omega_{0}/2$ and $a_{1}=\sqrt{\hbar / m \omega_{0}}$. Notice that the possibility of detecting the term depending upon the deformation parameter effect ($\delta T_{c}^{\alpha}$) requires
that, if $\delta T^{(0)}_{c}$  is the experimental error related to the measurement of the condensation temperature when $\alpha=0$, then $\delta T^{(0)}_{c}<|\delta T_{c}^{\alpha}|$. In
our case this entails
\begin{eqnarray}
\label{exp0}
\delta T^{(0)}_{c}&<&\Bigg|\alpha\frac{(8 \pi
m)^{1/2}\zeta(\gamma-1/2)} {\zeta(\gamma)\gamma}
(R_{0}N)^{-\frac{1}{2\gamma}}
\\\nonumber&+&\alpha a \frac{4m\zeta(\gamma-1)}{\pi
\hbar\zeta(\gamma)\gamma}\ln
\Bigg(\frac{(R_{0}N)^{\frac{1}{\gamma}}}{m \alpha^{2}}\Bigg)\Bigg|.
\end{eqnarray}
For spherical traps $\gamma=3(s_{1}+2)/2s_{1}$. The shift in the condensation temperature caused by interactions is typically $5\times 10^{-2}$ , with a $1\%$ of error \cite{RP,X}, then from expression (\ref{exp0}) and the results given above, in the case of $^{39}_{19} K$, with a mass $15\times10^{-26}$ kg, $a\sim10^{-9}$m, and $\omega_{0}\sim 10$Hz, allows us to obtain a criterion on $|\delta T_{c}^{\alpha}|$ as a function of the number of particles when $|\xi_{1}|\lesssim 1$. For different values of the shape parameter $\gamma$ we obtain, $N>10^{33}$ for $s_{1}=1$, $N>10^{22}$ for $s_{1}=2$, $N>10^{17}$ for $s_{1}=4$, $N>10^{13}$ for $s_{1}=9$,  $N>10^{11}$ for $s_{1}=18$ and so on. Notice that expression (\ref{exp0}) is not valid for the case $s_{1}=6$, (which implies $\gamma=2$) due to the divergent behavior of $\zeta(1)$. This special case, $s_{1}=6$, defines a limit between a positive shift and a negative one, caused by the deformation parameter $\alpha$. In other words, the shift caused by the deformation parameter is positive when $s_{1}<6$, for a positive $\alpha$. Conversely, with $s_{1}>6$ the corresponding shift is negative. Notice that, if the parameter $s_{1}$ is sufficiently large then, the number of particles decreases, but the shift on the condensation temperature caused by the deformation parameter becomes negative.

On the other hand, keeping the number of particles fixed, with say $N \sim 10^{5}-10^{6}$, we obtain frequencies of order $\omega_{0}\sim 10^{-9}$Hz for $s_{1}=1$, $\omega_{0}\sim 10^{-10}$Hz for $s_{1}=2$, $\omega_{0}\sim 10^{-15}$Hz for $s_{1}=4$ and so on. In other words, for a fixed number of particles, $\omega_{0}\rightarrow 0$ if the parameter $s_{1}$ grows.

In the case of an anisotropic three-dimensional harmonic oscillator
potential $\gamma=3$, with $A_{l}=\hbar\omega_{l}/2$,
$a_{l}=\sqrt{\hbar/m\omega_{l}}$, then from expression (\ref{CRTI})
we obtain
\begin{eqnarray}
\label{CRTIOS} \frac{\Delta
T_{c}}{T_{0}}&\simeq&-\Bigl(\frac{a}{a_{ho}}\Bigr)\Bigg[\frac{2\zeta(3/2)\zeta(2)-G_{3/2}(1)}{(2\pi)^{1/2}3\zeta(3)^{5/6}}\Bigg]N^{1/6}
\\\nonumber&+&\alpha\frac{2^{3/2}\pi^{-1/2}\zeta(5/2)}{3\zeta(3)^{5/6}}\Bigl(\frac{m}{\hbar\bar{\omega}}\Bigr)^{1/2}N^{-1/6}
\\\nonumber&+& \alpha a \Bigl(\frac{4m\zeta(2)}{3\pi\hbar\zeta(3)}\Bigr)\ln \Bigg(\frac{2(\hbar \bar{\omega})N^{1/3}}{\zeta(3)m \alpha^{2}}\Bigg),
\end{eqnarray}
where we have used the usual definitions $\bar{\omega}=(\omega_{1}\omega_{2}\omega_{3})^{1/3}$ and $a_{ho}=\Bigl(\frac{\hbar}{m\bar{\omega}}\Bigr)^{1/2}$.
In this case, $N>10^{22}$ which corresponds to a shift on the condensation temperature of order $\sim 10^{-5}$, positive. Conversely, $\bar{\omega}\sim 10^{-10}$Hz, for a fixed $N\sim10^{5}-10^{6}$ corresponding to a shift on the condensation temperature of order $10^{-3}$.

It is noteworthy to mention that $\xi_{1}$ could be bounded up to $|\xi_{1}|\lesssim 10^3$ by using (\ref{CRTIOS}) for $N\sim 10^{5}-10^{6}$, and $\bar{\omega}\sim 10$Hz, that is, one order of magnitude less than the bound obtained in \cite{CastellanosClaus}, which is notable.
If we set $\alpha=0$ in expression (\ref{CRTIOS}) we recover the
result given in \cite{Giorgini}.

\section{Deformed Bosonic Gas and Finite Size Corrections}

In this section let us calculate the leading correction on the
condensation temperature caused by finite size effects in our
modified bosonic gas. The correction to the condensation temperature originates in the zero--point motion, or equivalently, at the associated ground--state energy $\epsilon_{0}$ \cite{Pethick,Jaouadi}.
Thus, at the condensation temperature the chemical potential is given by
$\mu=\epsilon_{0}$. Setting $\mu=\epsilon_{0}$  and $a=0$ in (\ref{NPINT1}), we obtain that the relative shift in the condensation temperature is given by
\begin{eqnarray}
 \frac{\Delta
T_{c}}{T_{0}}&=&-\epsilon_{0}\frac{\zeta(\gamma-1)}{\zeta(\gamma)}\Bigg(NV_{char}\Bigl(\frac{2\pi \hbar^{2}}{m}\Bigr)^{3/2}\Bigg)^{-1/\gamma}\\\nonumber &+&\alpha\Bigl(\frac{2m}{\pi}\Bigr)^{1/2}\frac{\zeta(\gamma-1/2)}{\zeta(\gamma)}\Bigg(NV_{char}\Bigl(\frac{2\pi \hbar^{2}}{m}\Bigr)^{3/2}\Bigg)^{-1/2\gamma}\\\nonumber&+& O(\epsilon_{0}^{2},\alpha^{2})\label{fin}.
\end{eqnarray}
For spherical traps, setting $\epsilon_{0}=c_{1}\hbar\omega_{0}$ \cite{Salasnich}, the shift on the condensation temperature can be expressed as follows
\begin{eqnarray}
\label{TCE}
\frac{\Delta T_{c}}{T_{0}}&\simeq&-c_{1}\hbar\omega_{0}\frac{\zeta(\gamma-1)}{\zeta(\gamma)}(\Omega_{s_{1}} N)^{-2s_{1}/3(s_{1}+2)}\\\nonumber &+&\alpha\Bigl(\frac{2m}{\pi}\Bigr)^{1/2}\frac{\zeta(\gamma-1/2)}{\zeta(\gamma)}(\Omega_{s_{1}} N)^{-s_{1}/3(s_{1}+2)}.
\end{eqnarray}
where
\begin{equation}
\Omega_{s_{1}}=V_{char_{s_{1}}}\Bigl(\frac{2\pi \hbar^{2}}{m}\Bigr)^{3/2}.
\end{equation}
For different values of the shape parameter $\gamma$, we obtain from expression (\ref{TCE}), for instance, in the case of linear traps $s_{1}=1$, $\frac{\Delta T_{c}}{T_{0}} \sim \epsilon_{0_{s_{1}=1}} N^{-2/9}+\alpha N^{-1/9}$. For $s_{1}=2$, which corresponds to an isotropic harmonic oscillator,  $\frac{\Delta T_{c}}{T_{0}} \sim \epsilon_{0_{s_{1}=2}} N^{-1/3}+\alpha N^{-1/6}$. For $s_{1}=3$,  $\frac{\Delta T_{c}}{T_{0}} \sim \epsilon_{0_{s_{1}=3}}N^{-2/5}+\alpha N^{-1/5}$.  For $s_{1}=4$,  $\frac{\Delta T_{c}}{T_{0}} \sim \epsilon_{0_{s_{1}=4}} N^{-4/9}+\alpha N^{-2/9}$, and so on.
The possibility of detecting the term depending upon the deformation parameter effect entails in this case
\begin{equation}
\delta T^{(0)}_{c}<\Bigg|\alpha\Bigl(\frac{2m}{\pi}\Bigr)^{1/2}\frac{\zeta(\gamma-1/2)}{\zeta(\gamma)}(\Omega_{s_{1}} N)^{-s_{1}/3(s_{1}+2)}\Bigg|
\label{exp}. \,\,\,\,\,\,\,\, 
\end{equation}
The shift in the condensation temperature caused by finite size effects is typically of order $10^{-2}$ \cite{Pethick}, then from expression (\ref{exp}) and the results given above, in the case of $^{39}_{19} K$ and $\omega_{0}\approx 10$Hz, leads to $N>10^{33}$ for $s_{1}=1$, $N>10^{17}$ for $s_{1}=4$, $N>10^{14}$ for $s_{1}=6$, $N>10^{13}$  for  $s_{1}=9$, and so on. Conversely, keeping the number of particles fixed with, let say $N\sim10^{3}-10^{6}$, we obtain, $\omega_{0}\sim 1.62\times10^{-9}$Hz for $s_{1}=1$, $\omega_{0}\sim 8.70\times10^{-10}$Hz for $s_{1}=2$, $\omega_{0}\sim 5.70 \times 10^{-10}$Hz for $s_{1}=4$ and so on. Here the parameter $s_{1}$ has the same behavior as in the interacting case, that is, $\omega_{0}\rightarrow 0$ implies large values for $s_{1}$, when the number of particles is fixed.  

For an anisotropic three-dimensional harmonic oscillator potential
($\gamma=3$), we obtain that the relative correction in the condensation temperature
is given by
\begin{eqnarray}
\label{shiftfinite} \frac{\Delta
T_{c}}{T_{0}}&=&-\frac{\zeta(2)}{3\zeta(3)^{2/3}}\frac{\epsilon_{0}}
{\hbar\bar{\omega}}N^{-1/3}
\\\nonumber&+&\alpha \frac{\zeta(5/2)}{3\zeta(3)^{5/6}}\Bigl(\frac{8m}{\pi\hbar\bar{\omega}}\Bigr)^{1/2}N^{-1/6}+O(\epsilon_{0}^{2},\alpha^{2}).
\end{eqnarray}
In this case, $N> 10^{22}$ which corresponds to a shift on the condensation temperature of order $\sim10^{-10}$. Additionally, keeping $N\sim 10^{3}-10^{6}$ implies $\bar{\omega}\sim 8.70\times10^{-10}$Hz, corresponding to a shift of order $10^{-8}$. If we set $\alpha=0$ then, we recover the
usual result \cite{grossmann,ketterle,Haugerud,Pethick}.

\section{Conclusions}

Using the formalism of the semiclassical approximation, we have
analyzed the Bose--Einstein condensation for a modified bosonic gas
trapped in a 3--D power law potential in three regimes, namely,
the thermodynamic limit, finite size systems, and weakly interacting systems.
We have deduced the shift on the condensation temperature in the
thermodynamic limit, in a weakly interacting systems, and finite size systems as well,
in function of the number of particles and the trap parameters, which are
valid for any potential defined by the generic 3--dimensional
power--law potential (\ref{potgen}) within the semiclassical
approximation. We have obtained a bound up to $|\xi_{1}|\lesssim 10^6$ for linear traps to $|\xi_{1}|\lesssim 10^2$ corresponding to a free gas in a box, and  $|\xi_{1}|\lesssim 10^4$ for harmonic oscillator type potential in the ideal case, under typical conditions. We stress here that an improvement of the precision in the condensation temperature measurement would also allow to improve the bounds on $\xi_{1}$. 

For weakly interacting systems, we have obtained for the case $|\xi_{1}|\lesssim 1$, that if the trap parameter $s_{1}$ is sufficiently large then, this decreases the number of particles, but lead to corrections on the condensation temperature of order $10^{-5}$ for any trap parameter $s_{1}$, which is approximately 3 orders of magnitude smaller than the typical correction $10 ^{-2}$.  Conversely, keeping the number of particles $N\sim 10^{5}-10^{6}$ fixed, leading to frequencies of order $10^{-15}-10^{-9}$ corresponding to shifts on the condensation temperature up to $10^{-5}-10^{-3}$, for different values of the shape parameter $s_{1}$. These problems could be solved, in principle, just tuning the interaction coupling by Fesh-bach resonances to very small values of the scattering length $a$, that is, almost to the ideal case, and then, reducing the contribution of interactions on the condensation temperature below the Planck-scale induced shift for a sufficient large parameter $s_{1}$. Nevertheless, these facts could affect the thermodynamical equilibrium of the system, involving some technical difficulties.  

On the other hand, finite size effects for sufficient large $s_{1}$ leads to a very small correction on the condensation temperature of order $10^{-10}$, for any trap parameter $s_{1}$ and fixed frequency $\omega_{0}$, which is 8 orders of magnitude smaller than the typical correction $10^{-2}$. It is a condition impossible to fulfill. Conversely, keeping the number of particles $N\sim 10^{5}-10^{6}$ fixed, lead to frequencies up to $10^{9}-10^{-15}$ corresponding to shifts on the condensation temperature of order $10^{-8}-10^{-5}$, for different values of the shape parameter $s_{1}$. In other words, these facts suggest that finite size effects are technologically impossible to be tuned below Planck-scale induced effects, at least for current experiments.

For fixed frequencies of order $10$Hz, in the case of a harmonic oscillator potential, we obtain $N>10^{22}$, which implies a shift on the condensation temperature of order $10^{-5}$, in a weakly interacting system, and a shift on the condensation temperature of order $10^{-10}$, in finite size systems. Conversely, for a fixed number of particles $N\sim10^{5}-10^{6}$, this leads to a shift of order $10^{-3}$ for weakly interacting systems, with $\bar{\omega}\sim 10^{-10}$. For finite size systems, $\omega \sim 10^{-8}$ corresponding to shifts of order $10^{-8}$. Notice that the relevant contributions coming from the product $(N \omega^{n})^{m}$, where $m$ and $n$, depend on the properties of the trap in question. These facts suggest that many--body contributions on the relevant thermodynamic functions associated with the condensate could be used, in principle, to constrain significantly the parameter $\xi_{1}$, which in our case, for instance, could be bounded up to $|\xi_{1}|\lesssim 10^3$ by using (\ref{CRTIOS}), under typical conditions. 
\bigskip

Here is important to emphasize that the possibility of a systematic error in the measurements due to the variation in the corresponding trap frequency, could affect the usual predictions on the corresponding shift in the condensation temperature. However, these systematic errors can be estimated to be less than $1\%$ \cite{RP,X}, or even less than $0.5\%$, as it was reported in \cite{thesis}. 
In fact, in reference \cite{RP} each measurement at a
given s--wave scattering, is compared with a reference measurement for small values of this parameter of order 
$ \sim 0.005$, with the same frequency $\omega$ and an approximately equal number of particles. Thus, under these circumstances, the relative shift $\Delta T_{c}/T_{0}$ depends only on the s--wave scattering length effects, that is,  is assumed to be unaffected from all independent effects, including systematic errors in the absolute calibration of N and finite-size effects. 
In addition, systematic errors are often more easily controlled at lower temperatures.

On the other hand, the relative shift $\Delta T_{c}/T_{0}$ caused by
interactions, is highly trap--dependent, as can be seen from expression (\ref{CRTI}). For instance, in the case of harmonic traps, long--range fluctuations are suppressed \cite{yukalov} and
the leading term in the relative shift $\Delta T_{c}/T_{0}$ can be calculated with perturbative methods \cite{ST} like in the present report. However, higher orders in the relative shift $%
\Delta T_{c}/T_{0}$, calculated by using a non--perturbative approach behave
as $b_{1}\delta ^{\prime }+(b_{2}^{\prime }\ln \delta ^{\prime
}+b_{2}^{\prime \prime })\delta ^{\prime 2}$, in the case of a harmonic oscillator potential, where $\delta ^{\prime
}\equiv a/\lambda $ with $\lambda $ the thermal de Broglie wavelength \cite{AR}.  Here a good fit \cite{yukalov} yields $b_{1}\simeq -3.426$, $%
b_{2}^{\prime }\simeq -45.86$ and $b_{2}^{\prime \prime }\simeq -155.0$. Thus, it could be interesting to look at the corrections in the relative shift on the condensation temperature caused by the deformation parameters by using a non--perturbative approach.

Finally, we must add that the possible detection of these corrections, could be out of the current technology. Nevertheless, it is remarkable that an adequate choice of the shape associated with the potential under consideration, together with the many--body contributions, open the possibility of planning specific scenarios that could be used, in principle, to obtain a possible measure of the effects caused by the quantum structure of space--time.

\begin{acknowledgments}
This research was supported by the Deutscher Akademischer
Austauschdienst (DAAD), under grant $A/09/77687$.\\

Dedicated to the loving memory of my father El\'ias Castellanos de Luna.
\end{acknowledgments}

\end{document}